\documentclass[aps,pre,
superscriptaddress,groupedaddress,showpacs,twocolumn,preprint,10pt]{revtex4-1}
\usepackage{graphicx,epsf}
\usepackage{xcolor}
\usepackage{psfrag}
\usepackage{hyperref}
\usepackage[english]{babel}
\usepackage{titlesec}

\begin{document}
\title{Universal size ratios {{of Gaussian polymers with complex architecture:}}\\ Radius of gyration vs hydrodynamic radius}

\author{Khristine Haidukivska}
\author{Viktoria Blavatska}
\affiliation{Institute for Condensed Matter Physics of the National Academy of Sciences of Ukraine, 1, Svientsitskii Str., 79011 Lviv, Ukraine, 
}
\author{Jaros{\l}aw Paturej}
\affiliation{Institute of Physics, University of Silesia, 75 Pu\l{}ku Piechoty 1, 41-500 Chorz{\'o}w, Poland}
\affiliation{Leibniz Institute of Polymer Research Dresden e.V., Hohe Str. 6, 01069 Dresden, Germany}
\email[Correspondence to~]{jaroslaw.paturej@us.edu.pl}

\begin{abstract}
	{ The present research is dedicated to provide deeper understanding of the impact of complex architecture of branched polymers on their
	 behaviour in solvents. The folding dynamics of macromolecules and hydrodynamics of polymer fluids are
strongly dependent on size and shape measures of  
	  single macromolecules, which in turn are determined by their topology.
For this aim,} we use combination of  analytical theory, based on path integration method, and molecular dynamics simulations to study
 structural  properties of complex Gaussian polymers containing $f^c$ linear branches and $f^r$ closed loops grafted to the central core.
Using theory we determine  the size measures such as gyration radius $R_g$ and the hydrodynamic radii $R_H$, 
and obtain the estimates for the size ratio
$R_g /R_H$ with its dependence on the functionality $f=f^c+f^r$ of grafted polymers.  
{ In particular, we obtain the quantitative estimate of compactification (decrease of size measure) of such complex polymer 
	architectures with increasing number of closed loops $f^r$ as compared with
	linear or star-shape molecules of the same total molecular weight.}  Numerical simulations corroborate theoretical prediction that $R_g /R_H$ decreases towards unity with increasing $f$.  
These findings provide qualitative description of complex polymers with different arm architecture in $\theta$ solutions.
\end{abstract}
\pacs{36.20.-r, 36.20.Ey, 64.60.ae}
\date{\today}
\maketitle

\section{Introduction}

{{Polymer macromolecules of complex branched structure attract considerable
attention both from academical \cite{S,Ferber97} and applied \cite{Gao04,Jeon18} perspective,  being encountered as building blocks of materials like synthetic and biological
gels \cite{gels}, thermoplastics \cite{bates}, melts and elastomers \cite{paturej1,paturej2}. High functionality of polymers provides novel properties with
applications in diverse fields like drug delivery  \cite{Li16}, tissue engineering
 \cite{Lee01}, super-soft materials \cite{daniel}, and  antibacterial surfaces \cite{Zhou10} etc. On the other hand,
multiple loop formation in macromolecules is often encountered and plays an important role in biological
processes such as stabilization of globular proteins \cite{Nagi97} or transcriptional regularization of
genes \cite{Towles09}.
In this concern, it is of fundamental interests to study conformational properties of  complex polymer architectures.
}}

In statistical description of  polymers, a considerable attention is paid
 to the universal quantities describing equilibrium size and shape  of  typical conformation
 adapted by individual macromolecule in a solvent \cite{Clo,Gennes}.
In particular,  many physical properties are manifestations of the underlaying polymer conformation, including
the hydrodynamic properties  of polymer fluids \cite{Torre01}, the folding dynamics and catalytic activity of proteins \cite{Quyang08} etc.
As a size measure of a single macromolecule  one usually considers  the mean square radius of gyration $R_g^2$, which is directly measurable in static scattering experiments \cite{Ferri01,Smilgies15}.
 Denoting coordinates of the monomers along the polymer chain by $\vec{r}_n$, $n = 1, \ldots,N$, this quantity is defined  as:
\begin{equation}
\langle R_g^2 \rangle = \frac{1}{2N^2} \sum_{n, m}\langle (\vec{r}_n-\vec{r}_m)^2 \rangle,\label{Rg}
\end{equation}
and is thus given by a trace of gyration tensor $\bf{Q}$ \cite{Aronovitz86}.
Here and below, $\langle (\ldots ) \rangle$ denotes ensemble average over possible polymer conformations. 
Another important quantity that characterizes  the size of a polymer coil  is hydrodynamic radius $R_H$, which is directly obtained in dynamic light scattering experiments \cite{Schmidt81,Varma84,Linegar10}.
This quantity was introduced based on the
following motivation \cite{Doi}.
According to  the Stokes-Einstein equation, the diffusion coefficient $D$ of a spherical particle of radius $R_s$  in a solvent of viscosity $\eta$ at temperature $T$ is given by:
\begin{equation}
D=\frac{k_BT}{6\pi\eta R_s} \label{Stok}
\end{equation}
 where $k_B$ is Boltzmann constant. In order to  generalize the above  relation for the case of molecules of more complex shape, their center-of-mass diffusion coefficient $D$
is given by  Eq.~(\ref{Stok}) with $R_s$
replaced by 
$R_H$. The latter is given as the average
of the reciprocal  distances between  all pairs of monomers  \cite{TERAOKA}:
\begin{equation}
\langle R_H^{-1} \rangle = \frac{1}{N^2} \sum_{n, m} \left\langle \frac{1} { |\vec{r}_n-\vec{r}_m|} \right\rangle. \label{Rh}
\end{equation}
Namely, $R_H$  is related with the averaged components of the Oseen tensor ${\bf H}_{nm}$ characterizing the hydrodynamic interactions between monomers $n$ and $m$ \cite{Kirkwood54}.
To compare $R^2_g$ and $R^{-1}_H$, it is convenient to introduce the universal size ratio
\begin{equation}
\rho=\sqrt{ R_g^2} / R_H , \label{ratio}
\end{equation}
which does not depend on any details of chemical microstructure and is governed by polymer architecture. In the present paper we  restrict our consideration to the
ideal (Gaussian) polymers, i.e. monomers have no excluded volume. {{This to a certain extent corresponds to the behavior of flexible polymers in the so-called $\theta$-solvents. Note that our theoretical approach is not capable to correctly capture structural properties of more rigid branched polymers like dendrimers
or molecular bottlebrushes. The  rigidity of these macromolecules is controlled by steric repulsions between connected branches or grafts}}.
This approach allows to obtain the  exact analytical results for the set of universal quantities characterizing conformational
properties of macromolecules.   In particular, for a linear Gaussian polymer chain   the exact analytical result for the ratio (\ref{ratio})
in $d=3$ dimensions
reads \cite{zimm,burchard,dunweg}:
\begin{equation}
 \rho_{{\rm chain}}= \frac8{3\sqrt{\pi}}\approx 1.5045.
 \label{ratiochain}
 \end{equation}
The universal ratio of a Gaussian ring polymer was calculated in Refs.~\cite{burchard,fukatsu,Uehara2016} and is given by
\begin{equation}
\rho_{{\rm ring}} = \frac{\sqrt{2\pi}}2\approx 1.2533.
\label{ratioring}
\end{equation}
The validity of theoretically derived ratios $\rho_{{\rm chain}}$ and  $\rho_{{\rm ring}}$ was confirmed in several simulation studies~\cite{dunweg,Uehara2016,Clisby16}.
\begin{figure}[t!]
	\begin{center}
		\includegraphics[scale=0.25]{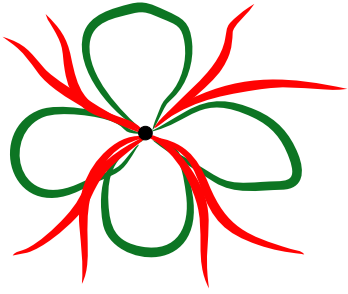}
		\caption{ \label{fig:1} Schematic presentation of rosette polymer topology comprised $f^r=4$ rings (green) and $f^c=8$ linear chains (red) grafted to a central core (black).}
\end{center}\end{figure}

\begin{table}[b!]
	\begin{tabular}{| c | c | c | c | c |}
		\hline
		Topology & $f^c$ &  $f^r$ & $\rho_{\mbox{\tiny theory}}$ & $\rho_{\mbox{\tiny sim}}$\\ \hline
		Chain & $1$ &  $0$ & $1.5045$ Eq.~(\ref{ratiochain}) & $1.5045\pm0.0005$ \cite{Clisby16} \\ \hline
		Ring & $0$ &  $1$ & $1.253$ Eq.~(\ref{ratioring}) & $1.253\pm0.013$ \cite{Uehara2016} \\ \hline
        Star & $3$ &  $0$ & $1.40$ Eq.~(\ref{ratiostar}) & $1.11$ \cite{Shida04}\\ \hline
        Star & $4$ &  $0$ & $1.33$ Eq.~(\ref{ratiostar})& $1.04$ \cite{Shida04}\\ \hline
        Tadpol& $1$ &  $1$ & $1.415$ Eq.~(\ref{ratiodring}) & $1.380\pm0.021$ \cite{Uehara2016} \\ \hline
		Double ring & $0$ &  $2$ & $1.217$ Eq.~(\ref{rhotadpole})  & $1.215\pm0.011$ \cite{Uehara2016} \\ \hline\end{tabular}
	\caption{Literature data for the universal size ratio for different polymer topologies, derived
using analytical theory $\rho_{\mbox{\tiny theory}}$ and numerical simulations $\rho_{\mbox{\tiny sim}}$.
The theoretical values for tadpol and double ring architectures  were calculated on the basis of our general analytical result, cf. Eq. (\ref{rhorosette}).
}\label{table1}
\end{table}

 The  distinct example of branched macromolecule is the so-called rosette  polymer \cite{Blavatska15}, containing  $f^c$ linear chains and $f^r$ closed loops (rings), radiating from the same branching point (see Fig. 1).
Note that for $f^r=0$ one restores architecture of a star polymer with $f^c$  functionalized linear chains radiating from a central core, for which  an exact analytical result is known for the size ratio (Ref. \cite{TERAOKA}):
\begin{equation}
\rho_{{\rm star}}=\frac{8\sqrt{f(3f^c-2)}}{3(f^c)^2\sqrt{\pi}}(\sqrt{2}-1)(\sqrt{2}+f^c). \label{ratiostar}
\end{equation}
The estimates for $\rho_{{\rm star}}$ have been also obtained  by numerical Monte-Carlo simulations \cite{Shida04}.
Using molecular dynamics (MD) simulations, Uehara and  Deguchi derived the universal size ratios for  macromolecules
 such as single ring ($f^c=0$, $f^r=1$), tadpole ($f^r=1$, $f^c=1$) and double ring ($f^r=2$, $f^c=0$) \cite{Uehara2016}. The
overview of existing literature data
for universal size ratios obtained in analytical $\rho_{\mbox{\tiny theory}}$ and numerical $\rho_{\mbox{\tiny sim}}$ investigations
are listed in
Table~\ref{table1}. Note large discrepancy between previous numerical study of star polymers \cite{Shida04}
and the theoretical result of Eq.~(\ref{ratiostar}). This significant difference between theory and simulations is due to
too short chains that were used in Ref.~\cite{Shida04} with maximum degree of polymerization $N=150$.
 As it will be shown the finite-size effect of polymer chains
strongly affects  measured value of $\rho$. In our numerical study we calculate $\rho$ in the asymptotic limit. For this purpose we simulated
long  polymer chains with degree of polymerization equal to $N=6400$.

The aim of the present work is to extend the previous analysis of rosette-like polymers \cite{Blavatska15},   by thoroughly studying their universal size characteristics. For this purpose we apply
the analytical theory, based on path-integration method, and extensive numerical molecular dynamics simulations. The layout of the paper is as follows.
In the next section, we introduce the continuous chain model and provide the details of analytical calculation of the universal size ratios $\rho$ for various
polymer architectures applying path integration method.
In section \ref{sec:MD} we describe the numerical model and details of MD simulations. In the same section we present numerical results and  compare them with our theoretical predictions.
We draw conclusions and remarks in section \ref{Con}.

\section{Analytical approach}\label{An}

\subsection{The model}

Within the frame of continuous chain model \cite{Edwards}, a single Gaussian polymer chain of length $L$ is represented as a path  $\vec{r}(s)$, parameterized by $0<s<L$.
We adapt this model to more complicated branched polymer topologies, containing in general  $f^c$ linear branches and $f^r$ closed rings (see figure \ref{fig:1}).
In the following, let us use notation $f=f^c+f^r$ for total functionality of such structure.
The weight of each $i$th path ($i=1,\ldots,f$) is given by
\begin{equation}
 W_i={\rm e}^{-\frac{1}{2}\int\limits_0^L ds \left(\frac{{\rm d}\vec{r}_i}{{\rm d}s}\right)^2}.
\end{equation}
The corresponding  partition function of rosette polymer is thus:
\begin{equation}
Z_{f^c,f^r} = \frac{\int\!{\cal D}\{\vec{r}\} \prod\limits_{j=1}^{f^r} \delta(\vec{r}_j(L){-}\vec{r}_j(0))\prod\limits_{i=1}^{f} \delta(\vec{r}_i(0))\, W_i}
{\int\!{\cal D}\{\vec{r}\} \prod\limits_{i=1}^{f} \delta(\vec{r}_i(0)) \, W_i},\label{Z}
\end{equation}
where ${\cal D}\,\{\vec{r}\}$ denotes multiple path integration over trajectories $\vec{r}_i(s)$ ($i=1,\ldots,f$) assumed to be of equal length
 $L_i=L$, the first product of $\delta$-functions reflects the fact that all $f^c+f^r$ trajectories start
at the same point (central core),  and the second $\delta$-functions product up to $f^r$  describes the closed ring structures of  $f^r$ trajectories (their starting and end points coincide). Note that (\ref{Z}) is  normalised in such a way that
the partition function of the system consisting of $f^c+f^r$ open linear Gaussian chains (star-like structure) is unity.
The expression for partition function of rosette-like polymer architecture have been evaluated in {{Ref.~\cite{Blavatska15}
and in Gaussian approximation reads:}}
\begin{equation}
Z_{f^c,f^r} =(2\pi L)^{-df^r/2}.
\end{equation}
{{where $d$ denotes spatial dimensionality.}} 
Within the frame of presented model, the expression for the mean square gyration radius from Eq.~(\ref{Rg}) can be rewritten as
\begin{equation}
\langle R_g^2 \rangle = \frac{1}{2(fL)^2} \sum_{i,j=1}^{f}\int_0^L\int_0^{L}\,ds_2\,ds_1 \langle (\vec{r}_i(s_2)-\vec{r}_j(s_1))^2 \rangle,
\end{equation}
whereas  the expression (\ref{Rh}) for  hydrodynamic radius reads:
\begin{equation}
\langle R_H^{-1} \rangle = \frac{1}{(fL)^2} \sum_{i,j=1}^{f} \int_0^L\int_0^{L}\,ds_2\,ds_1 \langle |\vec{r}_i(s_2)-\vec{r}_j(s_1)|^{-1} \rangle, \label{rhc}
\end{equation}
where $\langle (\ldots) \rangle$ denotes averaging over an ensemble of all possible configurations defined as:
\begin{eqnarray}
&&\langle (\ldots) \rangle = \frac{1}{Z_{f_c,f_r}} \times\label{av}\\
&& \times \frac{\int\!{\cal D}\{\vec{r}\} \prod\limits_{j=1}^{f^r} \delta(\vec{r}_j(L){-}\vec{r}_j(0))\prod\limits_{i=1}^{f} \delta(\vec{r}_i(0))(\ldots\,) W_i}
{\int\!{\cal D}\{\vec{r}\} \prod\limits_{i=1}^{f} \delta(\vec{r}_i(0)) \, W_i}. \nonumber
\end{eqnarray}

\subsection{Calculation of hydrodynamic radius and universal size ratio}

The crucial point in the calculation of the hydrodynamic radius is utilization of  the following equality \cite{Haydukivska14}:
\begin{equation}
|\vec{r}|^{-1}{ =} (2\pi)^{-d}\!  \int {\rm d}\vec{k} \,2^{d-1} \pi^{\frac{d-1}{2}} \Gamma\left(\!\frac{d-1}{2}\!\right)  k^{1-d}{\rm e}^{i\vec{r}\vec{k}}.
\end{equation}
where $\Gamma(x)$ is Gamma function.
Applying the above expression to  Eq.~(\ref{rhc})  allows to rewrite the mean reciprocal distance from the definition of $R_H$ as
\begin{eqnarray}
&&\langle|\vec{r}_i(s_2)-\vec{r}_j(s_1)|^{-1}\rangle { =} (2\pi)^{-d}\!   \int {\rm d}\vec{k}\, 2^{d-1} \pi^{\frac{d-1}{2}}\times \nonumber\\
&& \times \Gamma\left(\!\frac{d-1}{2}\!\right) \, k^{1-d} \langle  \xi(s_1,s_2) \rangle \label{defk}
\end{eqnarray}
with notation
\begin{equation}
\xi(s_1,s_2) \equiv {\rm e}^{i\vec{k}(\vec{r}_i(s_2)-\vec{r}_j(s_1))}.
\end{equation}
Below we will apply path integration approach to calculate the mean reciprocal distances.

\begin{figure}[t!]
	\begin{center}
		\includegraphics[width=90mm]{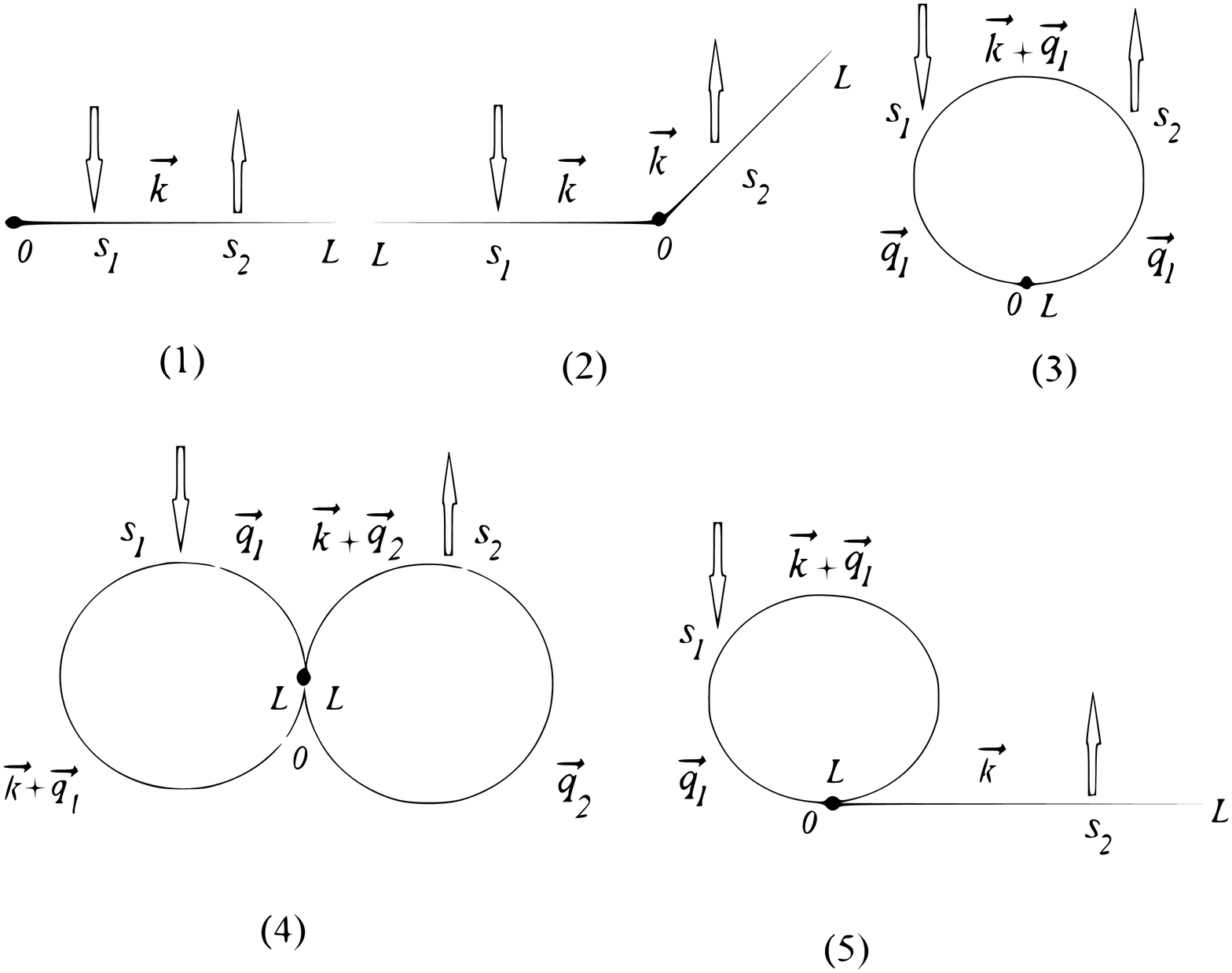}
		\caption{ \label{fig:2} Diagrammatic presentation of contributions into $\langle R_H^{-1} \rangle$ according to (\ref{rhc}). Solid lines are schematic presentation of polymer paths, arrows denote point $s_1$, $s_2$.   }
\end{center}\end{figure}

Exploiting the Fourier-transform of the  $\delta$-functions in definition (\ref{av})
\begin{equation}
\delta (\vec{r}_j(L)-\vec{r}_j(0)) =(2\pi)^{-d}\int {\rm d}\vec{q}_j\, {\rm e}^{-i\vec{q}_j(\vec{r}_j(L)-\vec{r}_j(0))} \label{d}
\end{equation}
we get a set of wave vectors $\vec{q}_j$ with $j=1,\ldots,f^r$ associated with $f^r$ closed loop trajectories, which is an important point in following evaluation.
To visualize different contributions into $\langle |\vec{r}_i(s_2)-\vec{r}_j(s_1)|^{-1}\rangle$,
it is convenient to use the diagrammatic technique (see Fig.~\ref{fig:2}).
Taking into account
the general rules of diagram calculations \cite{Clo}, each segment between any two restriction points $s_a$ and $s_b$
 is oriented and bears a wave vector $\vec{p}_{ab}$ given by a sum of incoming and outcoming  wave vectors injected
at restriction points and end points.
At these points, the flow of wave vectors is
conserved.
A factor $\exp\left(-{{p}_{ab}}^{\,\,2}(s_b-s_a)/2\right)$ is associated with each segment. An integration is to be made
over all independent segment areas and over wave vectors injected at the end points.

To make these rules more clear, let us start with  diagram (1), corresponding to the case when both  points $s_1$ and $s_2$ are
located along any linear arm of rosette polymer.  The vector $\vec{k}$ is injected at restriction point $s_1$
and  the segment $s_2-s_1$ is associated with factor $\exp\left(-{{k}}^{\,\,2}(s_2-s_1)/2\right)$.
Next step is performing integration over $k$. Passing to $d$-dimensional spherical coordinates, we have:
\begin{equation}
\int {\rm d}\vec{k}\, k^{1-d} f(k^2)= \frac{2\pi^{d/2}}{\Gamma\left( \frac{d}{2}\right)}\int{\rm d} {k}\, f(k^2),
\end{equation}
and thus integration over $k$ can be easily performed
\begin{equation}
\int_0^{\infty}\!{\rm d} k\, {\rm e}^{-\frac{k^2(s_2-s_1)}{2}}=\sqrt{  \frac{\pi}{2} }(s_2-s_1)^{-1/2}.
\end{equation}
The analytic expression corresponding to contribution from diagram (1) thus reads
\begin{equation}
\langle \xi(s_1,s_2)\rangle_{(1)} {=}\frac{(2\pi^{d+1})^{\frac{1}{2}}}{\Gamma\left(
\frac{d}{2}\right)}\!\int_0^L\!\!{\rm d}s_2\int_0^{s_2}\!\!\!{\rm d}s_1\left(s_2{-}s_1\right)^{-\frac{1}{2}}. \label{r1}
\end{equation}
Diagram (2) describes the situation when restriction points $s_1$ and $s_2$ are
located along two different linear arms of rosette polymer. We thus have a segment of length $(s_2+s_1)$ between them, associated with factor
$\exp\left(-{{k}}^{\,\,2}(s_2+s_1)/2\right)$. After  performing integration over $k$ we receive
\begin{equation}
\langle \xi(s_1,s_2)\rangle_{(2)} { =}\frac{(2\pi^{d+1})^{\frac{1}{2}}}{\Gamma\left(
	\frac{d}{2}\right)}\int_0^L\!\!{\rm d}s_2\int_0^{L}\!\!\!{\rm d}s_1\left(s_2{+}s_1\right)^{-\frac{1}{2}}.
\end{equation}

In the case (3), both $s_1$ and $s_2$ are located on the closed loop, let it be the loop with $j=1$. Here, we
need to take into account the wave vector $\vec{q}_1$, ``circulating'' along this loop, so that three segments should be taken into account
with lengths
$s_1$, $s_2-s_1$, and $L-s_2$, correspondingly, with associated factors $\exp\left(-{{q_1}}^{\,\,2}s_1/2\right)$,
$\exp\left(-(q_1+k)^{\,\,2}(s_2-s_1)/2\right)$, $\exp\left(-{{q}}^{\,\,2}(L-s_2)/2\right)$. Integration over the wave vector $q_1$ gives
\begin{eqnarray}
&&(2\pi)^{-d}\!\int{\rm d}\vec{q}_1\, {\rm e}^{-\frac{q_1^2L}{2}-\vec{q}\vec{k}(s_2-s_1)}=\nonumber\\
&&=(2\pi L)^{-d/2}(s_2-s_1)^{-\frac{1}{2}}{\rm e}^{\frac{k^2(s_2-s_1)^2}{2L}}.
\end{eqnarray}
After performing final integration over $k$ we receive
\begin{eqnarray}
&&\langle \xi(s_1,s_2)\rangle_{(3)}  {=}\frac{(2\pi^{d+1})^{\frac{1}{2}}}{\Gamma\left(
	\frac{d}{2}\right)}\int_0^L\!\!ds_2\int_0^{s_2}\!\!ds_1 \nonumber\\
&&\times\left(s_2{-}s_1-\frac{(s_2{-}s_1)^2}{L}\right)^{-\frac{1}{2}}.
 \end{eqnarray}
Following the same scheme, we receive  analytic expressions, corresponding to diagrams (4) and (5) on Fig.~\ref{fig:2}:
\begin{eqnarray}
&&\langle \xi(s_1,s_2)\rangle_{(4)}  {=}\frac{(2\pi^{d+1})^{\frac{1}{2}}}{\Gamma\left(
	\frac{d}{2}\right)}\int_0^L\!\!ds_2\int_0^{L}\!\!ds_1\nonumber\\
&&\times\left(s_2{+}s_1-  \frac{s_2^2}{L}{-}\frac{s_1^2}{L}\right)^{-\frac{1}{2}}, \\
&&\langle \xi(s_1,s_2)\rangle_{(5)}  {=}\frac{(2\pi^{d+1})^{\frac{1}{2}}}{\Gamma\left(
	\frac{d}{2}\right)}\int_0^L\!\!ds_2\int_0^{L}\!\!ds_1\nonumber\\
&&\times\left(s_2{+}s_1-\frac{s_1^2}{L}\right)^{-\frac{1}{2}}.\label{r5}
\end{eqnarray}
Note that  each diagram in Fig.~\ref{fig:2} is associated with the corresponding combinatorial factor.
 Namely, the contribution (1) in above expressions is taken with the pre-factor $f^c$, contribution (2)  with $\frac{f^c(f^c-1)}{2}$,  (3)  with   $f^r$, (4) with $\frac{f^r(f^r-1)}{2}$ and the last contribution (5) with the pre-factor $f^r f^c$.
Summing up all contributions from Eq. (\ref{r5})  with taking into account corresponding pre-factors,
on the base of Eq. (\ref{defk}) we finally obtain the expression for the hydrodynamic radius of a rosette structure:
\begin{eqnarray}
&&\langle R_{h,{\mbox{\tiny rosette}}}\rangle=\frac{\Gamma\left(\frac{d-1}{2}\right) }{\Gamma\left(\frac{d}{2}\right)\sqrt{2}}12(f^c+f^r)^2\sqrt{L}\times\nonumber\\
&&\left[-6 f_r\pi \left(\sqrt{2}(f^r-1)-2 f^r+1\right)+\right.\nonumber\\
&&16 \left( \sqrt {2}-1 \right) f^c \left( \sqrt {2}+ f^c \right) +\nonumber\\
&&\left.3f^cf^r \left( 10 \arcsin \left( \frac{\sqrt {5}}{5} \right) -\pi +4 \right)\right]^{-1}.\label{rha}
\end{eqnarray}

The expression for the mean square gyration radius of a rosette architecture is \cite{Blavatska15}:
\begin{eqnarray}
&&\langle R^2_{g,{\mbox{\tiny rosette}}}\rangle=\frac{Ld}{12(f^r+f^c)^2}[f^r(2f^r-1)+\nonumber\\
&&2f^c(3f^c-2)+8f^rf^c]. \label{rga}
\end{eqnarray}

Finally,  using Eqs.~(\ref{rha}) and (\ref{rga}), we calculate the  the universal size ratio (\ref{ratio}) of rosette-like polymer architecture in Gaussian approximation:
\begin{eqnarray}
&&\rho_{{\mbox{\tiny rosette}}}=\frac{\sqrt{6\,d}\,\Gamma\left(\frac{d-1}{2}\right)}{72(f^r+f^c)^3\Gamma\left(\frac{d}{2}\right)}\times\nonumber\\
&&\sqrt{6(f^c)^2+8f^cf^r+2(f^r)^2-4f^c-f^r}\times\nonumber\\
&&\left[-6 f_r\pi \left(\sqrt{2}(f^r-1)-2 f^r+1\right)+\right.\nonumber\\
&&16 \left( \sqrt {2}-1 \right) f^c \left( \sqrt {2}+ f^c \right) +\nonumber\\
&&\left.3f^cf^r \left( 10 \arcsin \left( \frac{\sqrt {5}}{5} \right) -\pi +4 \right)\right]. \label{rhorosette}
\end{eqnarray}
Substituting $d=3$ in  expression (\ref{rhorosette}),
 for $f^r=0$, { both at $f^c=1$ and $f^c=2$}  we restore the universal size ratio of  a linear polymer (\ref{ratiochain}),
whereas $f^c>2$ and $f^r=0$ gives the expression for a star polymer (\ref{ratiostar}).
 For $f^c=0$ and  $f^r=1$ we reproduce the known analytical expression of a single ring from Eq.~(\ref{ratioring}).
 Consequently $f^c=0$ and  $f^r=2$  Eq.~(\ref{rhorosette}) provides the formula for universal size ratio of a star comprised of
 two ring polymers:
\begin{equation}
\rho_{{\mbox{\tiny double ring}}}=\frac{\sqrt{3\pi}}{4}(3-\sqrt{2})\approx 1.217. \label{ratiodring}
\end{equation}
For $f^c=1$ and $f^r=1$ we find analytic expression for the so-called tadpole architecture:
 \begin{eqnarray}
&&\rho_{{\mbox{\tiny tadpole}}}=\frac{\sqrt{22}}{96\sqrt{\pi}}\left[3\pi+28+30\arcsin\left(\frac{\sqrt{5}}{5}\right)\right]\nonumber\\
&&\;\;\;\;\;\;\;\;\;\;\;\approx 1.415. \label{rhotadpole}
\end{eqnarray}

\begin{figure}[t!]
	\begin{center}
		\includegraphics[scale=0.3]{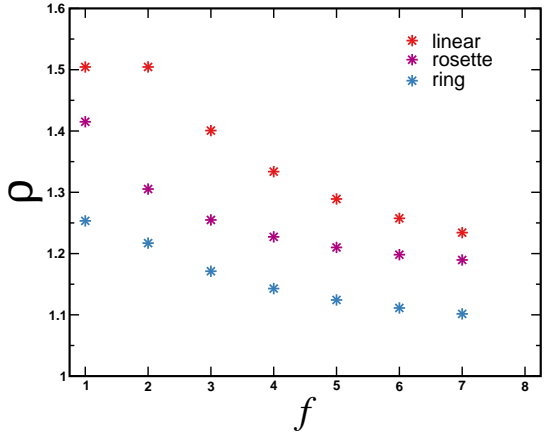}
		\caption{Summary of theoretical results for universal size ratio $\rho$  as given by (\ref{rhorosette})
vs functionality $f=f^c+f^r$
for different polymer topologies.
 Data for architectures containing: only linear chains (star-like polymer with $f^r=0$) as function of $f=f^c$ (red symbols),
 only ring polymer (with $f^c=0$) as function of $f=f^r$ (blue symbols) and
``symmetric'' rosette structure with equal number of rings and linear branches $f=f^r+f^c$ (purple symbols).}\label{theory}
\end{center}\end{figure}

In Fig.~\ref{theory} we plot calculated theoretical values of the universal size ratio vs number of functionalized chains  for stars comprised of linear polymers with $f^c>0$, $f^r=0$ (red symbols) and ring polymers $f^r>0$, $f^c=0$ (blue) as well as rosette polymers
with equal number of grafted linear chains and rings $f^r=f^c>0$ (purple).
For all architectures we observe  decrease in $\rho$ with increasing  functionality.
In the next subsection we compare our theoretical predictions with the result of MD simulations.

\section{Numerical approach}
\label{sec:MD}

\subsection{The method}

Numerical data in this work have been obtained from
 MD simulations.
 We consider simple three-dimensional numerical model of a rosette polymer consisting of arms which are $f^c$ linear chains and/or $f^r$ ring polymers.
  Each arm is composed of $N$ sizeless particles
of equal mass $m$ connected by bonds.
 We study ideal (Gaussian) conformations of rosette polymers {{corresponding to a certain extent to the conformations of real rosette polymers  at dilute $\theta$ solvent conditions.}}
  In our numerical model the connectivity along the polymer chain backbone is assured via harmonic potential
 \begin{equation}
 V(r)=\frac k2(r-r_0)^2,
\label{harmonic}
 \end{equation}
where  $k=200$~$k_BT/b^2$ is the interaction strength measured in units of thermal energy $k_BT$ and  and the equilibrium bond distance
 $r_0=b$.

The molecular dynamics  simulations were performed 
by solving the Langevin equation of motion for
the position $\vec{r}_i=[x_i,y_i,z_i]$ of each monomer,
\begin{equation}
m\ddot{\vec{r}}_i =   \vec{F}_i -\zeta\dot{\vec{r}}_i + \vec{F}_i^{\mbox{\tiny R}}, \,\,\,
i=1,\ldots,fN,
\label{langevin}
\end{equation}
which describes the motion of bonded monomers.
Forces $\vec{F}_i$ in Eq.~(\ref{langevin}) above
are obtained from the harmonic interaction potential between (Eq.~\ref{harmonic}).
The second and third term on the right hand side of Eq.~(\ref{langevin})
 is a slowly evolving viscous force $-\zeta\dot{\vec{r}}_i$
and a rapidly fluctuating stochastic force
$\vec{F}_i^{\mbox{\tiny R}}$ respectively.  This random force $\vec{F}_i^{\mbox{\tiny R}}$ is related to the friction coefficient $\zeta$ by the fluctuation-dissipation
theorem $\langle \vec{F}_i^{\mbox{\tiny R}}(t) \vec{F}_j^{\mbox{\tiny R}}(t')\rangle = k_BT\zeta \delta_{ij}\delta(t-t')$.
The friction coefficient used in simulations was $\zeta=0.5\,m\tau^{-1}$ where   $\tau = [mb^2/(k_BT)]^{1/2}$
is the unit of time.
 A Langevin thermostat was
used to keep the temperature constant. The integration step employed to solve the equations of motions was taken to be
$\Delta t=0.0025\tau$.
 All simulations were performed in a cubic box 
with periodic boundary conditions imposed in all spatial dimensions.
We used Large-scale Atomic/Molecular Massively Parallel Simulator
(LAMMPS) \cite{lammps} to perform simulations. Simulation snapshots were rendered using Visual
Molecular Dynamics (VMD) \cite{vmd}.
\begin{figure}[!h]
	\begin{center}
		\includegraphics[scale=0.22]{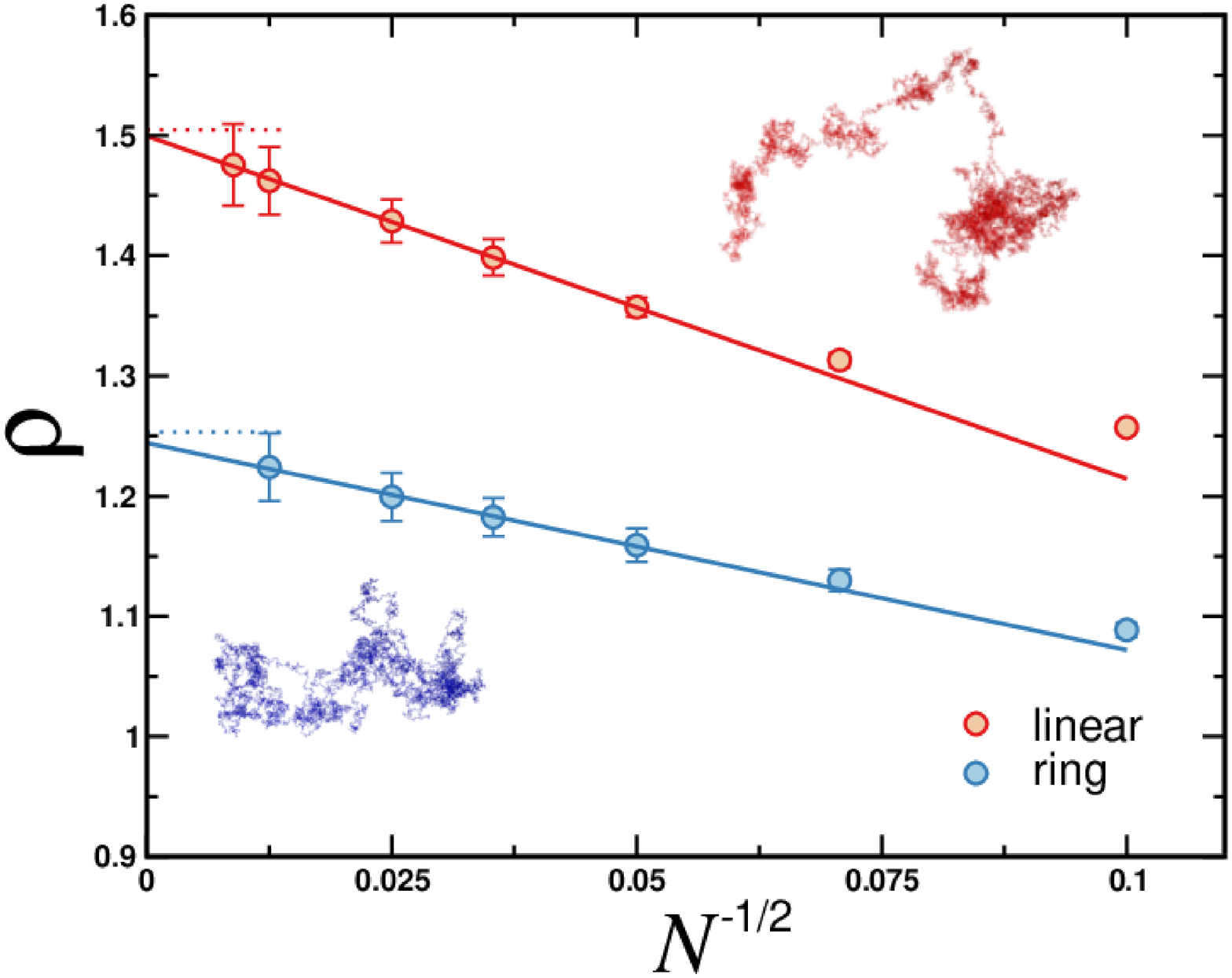}
		\caption{Molecular dynamics data for the universal size ratio $\rho$ of linear chains (red symbols) and ring polymers (blue symbols) plotted
as a function of correction-to-scaling variable $N^{-1/2}$ with corresponding simulation snapshots for polymer architectures with degree of polymerization $N=6400$. Solid lines represent fitting functions of the general form given in Eq.~(\ref{corr}). Horizontal dotted lines correspond to
asymptotic values $\rho_{\infty}$ predicted by theory, cf. Eqs.~(\ref{ratiochain}) and (\ref{ratioring}).
\label{MD1}
}
\end{center}
\end{figure}

\subsection{Results }
Simulations of  rosette polymers were performed for the following number of monomer beads per arm $N=100,200,400,800,1600$ and 6400.
The number of arms for star polymers composed of solely linear chains (i.e.~with $f^r$=0) and ring polymers (i.e.~with $f^c=0$) were varied in the range between 1 to 4. In the case of rosette polymers which are hybrid polymer architectures comprised of linear chains and ring polymers we considered two arm functionalities with $f^c=f^r=1$ and 2.
 To increase conformational sampling each simulation was carried out with 50 identical molecules in a simulation box.
In the course of simulations the universal size ratio was measured, cf.~Eq.~(\ref{ratio}).
 In the numerical calculation  of  quantities like $\rho$ a crucial aspect  is  finite degree of polymerization  $N$ that we are dealing with in simulations, while theoretically obtained values of $\rho$ hold in the asymptotic limit $N\to\infty$.
Thus, the finite-size effects (or corrections to scaling) should be appropriately taken into account.
For the size ratio of an ideal linear chain, this correction is given by
\begin{equation}
\rho=\rho_{\infty}(1+aN^{-\Delta}), \label{corr}
\end{equation}
where $\rho_{\infty}$ is the asymptotic value obtained at $N\to\infty$, $a$ is non-universal amplitude, $\Delta$ is the correction-to-scaling exponent for
$\theta$-solvent is $\Delta=1/2$ \cite{dunweg} whereas for good solvent conditions
is $\simeq0.53$ \cite{Clisby16}.
In our numerical analysis we use Eq.~(\ref{corr}) to obtain the universal size ratio in the asymptotic limit  for all considered architectures.
For this purpose we plot $\rho$ vs correction-to-scaling term $N^{-1/2}$ and get  $\rho=\rho_{\infty}$ for $N\rightarrow \infty$.

In Fig.~\ref{MD1} we display the results of our MD simulations for two "benchmark" systems which are Gaussian linear chain (red circles)
and  Gaussian ring (blue circles).
For both architectures systematic increase in the size ratio is observed with increasing value of $N$.
In the asymptotic limit $N\rightarrow \infty$  we obtain
$\rho_{\mbox{\tiny chain}}=1.499\pm0.005$ and $\rho_{\mbox{\tiny ring}}=1.244\pm0.004$. 
These numerical values with very good accuracy reproduce known theoretical results.
The latter are given by Eq.~(\ref{ratiochain}) for linear chains  and by (\ref{ratioring}) for rings.
The complete list of numerically derived  universal size ratios and their comparison to theoretical values can be found in
Table \ref{table2}.

\begin{figure}[!h]
	\begin{center}
		\includegraphics[scale=0.23]{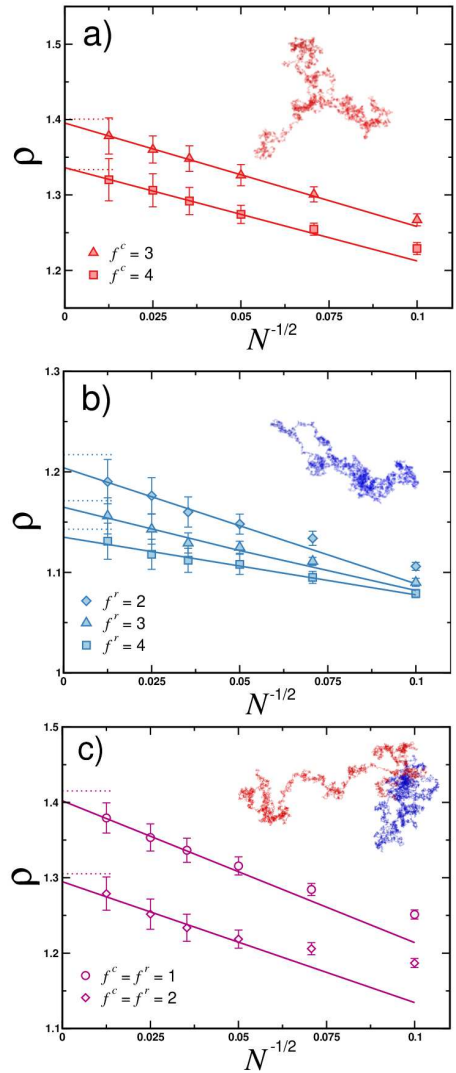}
		\caption{Molecular dynamics data for the universal size ratio $\rho$ of star polymers comprised of a) linear, b) ring polymers and c) rosette polymers
plotted
as a function of correction-to-scaling variable $N^{-1/2}$.
Data displayed for different amount of arms $f^c$ and $f^r$ as indicated. For rosette polymers data are for symmetric number of arms $f^c=f^r$.
 Solid lines represent  fitting functions according  to Eq.~(\ref{corr}). Horizontal dotted  lines correspond to  asymptotic
values obtained from analytical theory, see Table \ref{table2}. Insets show simulation snapshots for with $N=6400$ and: a) $f^c=3$, b) $f^r=2$ and c) $f^c=f^r=1$.
\label{MD2}
}
\end{center}\end{figure}
\begin{table}[!h]
\begin{tabular}{| c | c | c | c |}
   \hline
$f^c$ &  $f^r$ & $\rho_{\mbox{\tiny theory}}$ & $\rho_{\mbox{\tiny sim}}$ \\ \hline
$1$ &  $0$ & $1.504$ & $1.499\pm0.005$ \\ \hline
$2$ &  $0$ & $1.504$ & $1.499\pm0.005$ \\ \hline
$3$ &  $1$ & $1.401$ & $1.395\pm0.006$ \\ \hline
$4$ &  $0$ & $1.334$ & $1.336\pm0.006$ \\ \hline\hline
$0$ &  $1$ & $1.253$ & $1.244\pm0.004$ \\ \hline
$0$ &  $2$ & $1.217$ & $1.204\pm0.010$ \\ \hline
$0$ &  $3$ & $1.171$ & $1.165\pm0.011$ \\ \hline
$0$ &  $4$ & $1.143$ & $1.135\pm0.012$ \\ \hline\hline
$1$ &  $1$ & $1.415$ & $1.401\pm0.008$ \\ \hline
$2$ &  $2$ & $1.305$ & $1.295\pm0.018$ \\ \hline
\end{tabular}
\caption{Summary of  theoretical results for the size ratio $\rho_{\mbox{\tiny theory}}$ calculated using Eq.~(\ref{rhorosette}) and asymptotic values $\rho_{\mbox{\tiny sim}}$ obtained from MD simulations for rosette polymer architectures
comprised of different number of $f^c$ linear chains and $f^r$ ring polymers.
}\label{table2}
\end{table}

In Fig.~\ref{MD2} we show numerically derived universal size ratios as a function of $N^{-1/2}$ for more complex architectures.
We investigated conformations of  stars comprised  of linear chains, stars of ring polymers and
rosette polymers with equal number of grafted linear and ring chains. For all architectures we observe
systematic approaching to asymptotic values predicted by theory with increasing value of $N$ per arm.
 For
stars of linear chains with functionality $f^c=3$ and 4 (cf.~Fig.~\ref{MD2}a) simulations provide the following  universal size ratios:  $1.395\pm0.006$ and $1.336\pm0.006$. Both values are with very good agreement to the theoretical prediction
given by Eq.~(\ref{ratiostar}).
Note that the values of $\rho$ calculated in the course of our simulations are much closer to the  analytical theory results as compared to existing numerical data \cite{Shida04}.
For stars comprised of cyclic macromolecules (cf.~Fig.~\ref{MD2}b) we reproduce the theoretical value of Eq.~(\ref{ratiodring}) for double ring architecture ($f^r=2$)
 as well as for stars with larger number of grafted rings, cf. Eq.~(\ref{rhorosette}) with $f^c=0$ and $f^r=3$ or 4.
Namely, we get $1.204\pm0.010$ for $f^r=2$, $1.165\pm0.011$ for $f^r=3$ and $1.135\pm0.012$ for $f^r=4$.
For the tadpole architecture, the  simplest rosette polymer which is comprised of $f^c=1$ and $f^r=1$  arms (see snapshot in Fig.~\ref{MD2}c), we
obtain the size ratio of  $1.401\pm0.008$ which matches theoretically predicted value for this type of polymer from Eq.~(\ref{rhotadpole}).
For rosette polymers with $f^c=2$ and $f^r=2$ our simulations provide  $1.295\pm0.018$ which is comparable with the corresponding
value calculated from the formula given in Eq.~(\ref{rhorosette}). The full list of calculated values of $\rho$ is in Table \ref{table2}.



\section{Conclusions}\label{Con}

We have studied by  combination of analytical theory and molecular dynamics
simulations  conformational properties of rosette polymers which are complex macromolecules
consisting of  $f^c$ linear chains (branches) and $f^r$ closed loops (rings) radiating from the central branching point.
Our focus was on characterizing structure of ideal polymer conformation with no excluded volume interactions. For this purpose we investigated
 basic structural quantities
 such as the mean square radius of gyration $R_g^2$, the hydrodynamic radius $R^{-1}_H$ and most importantly  the universal size ratio $\rho \equiv\sqrt{R_g^2}/R_H$.
Our calculations demonstrated gradual decrease in $\rho$ with increasing functionality $f=f^c+f^r$  of grafted polymers.
The analytical results are in perfect agreement with our numerical simulations data. 
Since both quantities $R_g^2$ and $R_H$ are directly accessible via correspondingly static  and dynamic scattering techniques we hope
that our results will stimulate further experimental studies on the behavior of complex polymer architectures in solutions.

\begin{acknowledgements}
J.P. acknowledged the support from the National Science Center, Poland (Grant No. 2018/30/E/ST3/00428) and
 the computational time at PL-Grid,  Poland.
\end{acknowledgements}






\begin{thebibliography}{99}
	
\bibitem{S}
L. Sch\"afer L, C. von Ferber, U.  Lehr, and B.  Duplantier,   Nucl. Phys. B {\bf 374}, 473 (1992)

\bibitem{Ferber97}
C. von Ferber and Yu.  Holovatch,   Phys. Rev. E {\bf 56}, 6370 (1997)

\bibitem{Gao04}
C. Gao  and D. Yan,  Prog. Polym. Sci. {\bf 29}, 183 (2004)

\bibitem{Jeon18}
I.-Y. Jeon, H.J. Noh, and J.B. Baek,  Molecules {\bf 23}, 657
(2018)

\bibitem{gels} M. Djabourov, K. Nishinari, and S.B. Ross-Murphy,  {\it  Physical Gels from Biological and Synthetic Polymers} (Cambridge University Press, Cambridge, 2013)

\bibitem{bates} J. Zhang, D.K. Scheiderman, T. Li, M.A. Hillmyer, and F.S. Bates, {Macromolecules} {\bf 49}, 9108 (2016)


\bibitem{paturej1}  J. Paturej and T. Kreer,  { Soft Matter} {\bf 13}, 8534 (2017)

\bibitem{paturej2} J. Paturej, S. Sheiko, S. Panyukov and M. Rubinstein, { Science Advances} {\bf 2}, e1601478 (2016)

\bibitem{Li16}
J. Li  and D.J. Mooney, Nat. Rev. Mater. {\bf 1}, 16071 (2016)

\bibitem{Lee01}
K. Y. Lee  and D.J. Mooney,  Chem. Rev. {\bf 101}, 1869 (2001)


\bibitem{daniel} W. Daniel, J. Burdy\'nska, M. Vatankhah-Varnoosfaderani, K. Matyjaszewski, J. Paturej, M. Rubinstein,
A.V. Dobrynin and S.S. Sheiko,
{Nature Materials} {\bf 15}, 183 (2016)

\bibitem{Zhou10}
Y. Zhou, W. Huang, J. Liu, X. Zhu, and D. Yan,  Adv. Mater. {\bf 22}, 4567  (2010)


\bibitem{Nagi97}
A.D. Nagi and L. Regan, Folding Des. {\bf 2}, 67 (1997)

\bibitem{Towles09}
K. B. Towles, J.F. Beausang, H.G. Garcia, R. Phillips, and P.C. Nelson, Phys. Biol. {\bf 6}, 025001
(2009)


\bibitem{Clo}
J. Des Cloizeaux, G. Jannink, {\it  Polymers in Solution: Their Modeling and Structure} (Clarendon Press, Oxford, 1990)

\bibitem{Gennes}
P.G. de Gennes, {\it  Scaling Concepts in Polymer Physics} (Ithaca, NY: Cornell University Press 1979)

\bibitem{Torre01}
G. de la Torre, O. Llorca, J.L. Carrascosa, and J.M. Valpuesta, Eur. Biophys. J. {\bf 30}, 457 (2001)

\bibitem{Quyang08}
 Z. Quyang and J. Liang, Protein Sci. {\bf  17}, 1256 (2008)


\bibitem{Ferri01}
F. Ferri,  M. Greco, and   M. Rocco, Macromol. Symposia {\bf  162}  (2000) 23-44

\bibitem{Smilgies15}
D.-M. Smilgies and E. Folta-Stogniew, J. Appl. Crystallogr.  {\bf 48} 1604 (2015)

\bibitem{Aronovitz86}
J.A. Aronovitz and D.R. Nelson, J. Physique {\bf 47}, 1445 (1986);
J. Rudnick and G. Gaspari, J. Phys. A {\bf 19}, L191 (1986);
G. Gaspari, J. Rudnick, and A. Beldjenna, J. Phys. A {\bf 20}, 3393 (1987).


\bibitem{Schmidt81}
M. Schmidt and W. Burchard, Macromolecules  {\bf 14}, 210 (1981)

\bibitem{Varma84}
B.K. Varma, Y. Fujita, M. Takahashi, and T. Nose,
J.  Polym. Sci. Polym.  Phys. Ed. {\bf 22}, 1781 (1984)

\bibitem{Linegar10}
K.L. Linegar, A.E. Adeniran, A.F. Kostko, and M.A. Anisimov,  Colloid Journal {\bf 72}, 279 (2010)

\bibitem{Doi}
M. Doi and S. F. Edwards, {\it The Theory of Polymer Dynamics}
(Oxford University Press, Oxford,  1988).

\bibitem{TERAOKA}
I. Teraoka, {\it Polymer Solutions:
	An Introduction to Physical Properties},  (John Wiley \& Sons Inc, New York, 2002)


\bibitem{Kirkwood54}
J. G. Kirkwood, J. Polym. Sci. {\bf 12}, 1 (1953)

\bibitem{zimm} B. H. Zimm  and W.H.J. Stockmayer,  { J. Chem. Phys.} {\bf 17}, 1301 (1949)

\bibitem{burchard} W. Burchard and M. Schmidt, { Polymer} {\bf 21}, 745 (1980)

\bibitem{dunweg} B. D\"unweg, D. Reith, M. Steinhauser  and K. Kremer, { J. Chem. Phys.} {\bf 117}, 914 (2002)

\bibitem{fukatsu} M. Fukatsu, M.J. and Kurata, { J. Chem. Phys.} {\bf 44}, 4539 (1966)

\bibitem{Uehara2016}
E. Uehara and T. Deguchi,  J. Chem. Phys. {\bf 145}, 164905 (2016)

\bibitem{Clisby16}
N. Clisby and B. D\"unweg, Phys. Rev. E {\bf 94}, 052102 (2016)




\bibitem{Blavatska15}
V. Blavatska, R. Metzler,  J. Phys. A: Math. Theor. {\bf 48}, 135001  (2015)



\bibitem{Shida04}
K. Shida, K.  Ohno,  M.  Y. Kawazoe, and
Y. Nakamura, Polymer {\bf 45},  1729 (2004)


\bibitem{Edwards}
S.F. Edwards,  Proc. Phys. Soc. Lond. {\bf 85},  613  (1965);  Proc. Phys. Soc. Lond. {\bf 88},  265 (1965)

\bibitem{Haydukivska14}  K. Haydukivska and V. Blavatska, J. Chem. Phys. {\bf 141}, 094906 (2014)


\bibitem{lammps}  S.J. Plimpton, { J. Comp. Phys.} {\bf 117}, 1 (1995) (http://lammps.sandia.gov)

\bibitem{vmd}  W. Humphrey, A  Dalke,  and K. Schulten, { J. Mol. Graphics} {\bf 14}, 33 (1996)







\end{thebibliography}
\end{document}